# Single gap superconductivity in $\beta$-Bi$_2$Pd

03/02/2016

J. Kačmarčík,[1] Z. Pribulová,[1] T. Samuely,[1] P. Szabó,[1] V. Cambel,[2] J. Šoltýs,[2] E. Herrera,[3] H. Suderow,[3] A. Correa-Orellana,[4] D. Prabhakaran,[5] and P. Samuely[1*]

[1]*Centre of Low Temperature Physics @ Institute of Experimental Physics, Slovak Academy of Sciences and P. J. Šafárik University, 040 01 Košice, Slovakia*
[2]*Institute of Electrical Engineering, Slovak Academy of Sciences, Dubravska cesta 9, 84104 Bratislava, Slovakia*
[3]*Laboratorio de Bajas Temperaturas, Departamento de Física de la Materia Condensada, Instituto Nicolás Cabrera and Condensed Matter Physics Center, Universidad Autónoma de Madrid, E-28049 Madrid, Spain*
[4]*Instituto de Ciencia de Materiales de Madrid, Consejo Superior de Investigaciones Científicas (ICMM-CSIC), Sor Juana Inés de la Cruz 3, 28049 Madrid, Spain*
[5]*Department of Physics, Clarendon Laboratory, University of Oxford, Park Road, Oxford OX1 3PU, UK*

$\beta$-Bi$_2$Pd compound has been proposed as another example of a multi-gap superconductor [*Y. Imai et al., J. Phys. Soc. Jap.* **81**, *113708 (2012)*]. Here, we report on measurements of several important physical quantities capable to show a presence of multiple energy gaps on our superconducting single crystals of $\beta$-Bi$_2$Pd with the critical temperature $T_c$ close to 5 K. The calorimetric study via a sensitive ac technique shows a sharp anomaly at the superconducting transition, however only a single energy gap is detected. Also other characteristics inferred from calorimetric measurements as the field dependence of the Sommerfeld coefficient and the temperature and angular dependence of the upper critical magnetic field point unequivocally to standard single s-wave gap superconductivity. The Hall-probe magnetometry provides the same result from the analysis of the temperature dependence of the lower critical field. A single-gapped BCS density of states is detected by the scanning tunneling spectroscopy measurements. Then, the bulk as well as the surface sensitive probes evidence a standard conventional superconductivity in this system where the topologically protected surface states have been recently detected by ARPES [*M. Sakano et al., Nature Comm.* **6**, *8595 (2015)*].

## I. INTRODUCTION

The multigap superconductivity attracts a lot of attention in particular after its spectacular appearance in MgB$_2$ [1]. More recently other aspects of the phenomenon have been intensively studied in iron pnictides and selenides [2]. Recent theoretical studies forecast for the multigap systems new interesting physics such as the formation of the chiral ground state and the appearance of different spatial length scales of the band condensates which may result in a plethora of new phenomena like fractional vortices [3], flux-carrying topological solitons [4] and other exotic states [5]. Before any experimental exploration of these phenomena in a particular system its multigap nature must be satisfactorily confirmed. In MgB$_2$ a presence of two superconducting energy gaps has been strongly revealed in manifold characteristics. Indeed two gaps can be directly detected via spectroscopy sensitive to the gapped density of states like (Andreev reflection) tunneling [6] or ARPES [7]. In heat capacity the small gap on the MgB$_2$ $\pi$-band causes a strong low temperature hump while the large one of the $\sigma$-band give rise to a typical jump due to the second order phase transition [8]. Further, the magnetic field dependence of the Sommerfeld coefficient shows a strong nonlinearity in contrast to almost linear behavior in the case of a standard situation. The lower and upper critical fields behave also unusually leading for example to different superconducting anisotropies depending whether defined via $H_{c1}$ or $H_{c2}$, respectively [9]. Moreover such anisotropies are temperature dependent due to interplay of the gaps which is varying in different parts of the Field vs. Temperature phase diagram [10,11].

The multiband/multigap superconductivity has been proposed for the layered superconductor of $\beta$-Bi$_2$Pd with $T_c$ close to 5 K [12] because of its heat-capacity characteristics as well as the temperature dependence of the upper critical magnetic fields. Also the theoretical studies of the electronic band structure suggest a possibility of a complicated superconducting energy gap with different amplitudes on different sheets of the Fermi surface [13]. We have addressed the issue of multigap superconductivity by multiple techniques capable to provide the above mentioned characteristics sensitive to presence of multiple superconducting gaps. Particular measurements were always performed on several single crystals of $\beta$-Bi$_2$Pd. As a result we present the temperature dependence of the electronic part of the heat capacity, the magnetic field dependence of the Sommerfeld coefficient, the temperature dependence of the upper critical magnetic fields for the both principal field orientations, and the angular dependence of the critical temperature at a constant magnetic field, all inferred from a highly sensitive ac calorimetry. Further, from the Hall-probe magnetometry the temperature dependence of the lower critical field has been deduced. All these characteristics are fully compatible with the single s-wave gap superconductivity and no multiband effects are observed. The subkelvin scanning tunneling spectroscopy proved only a single superconducting gap, too. Thus, a conventional superconductivity has been proved for both bulk and surface of $\beta$-Bi$_2$Pd which is noteworthy in the system where topologically protected surface states had been reported [14]. Yet some characteristics as the upper critical magnetic field inferred from the magnetoresistive transitions and the superconducting coupling ratio from STS deviate from those obtained from the bulk probes indicating that some surface properties may differ from the bulk.

## II. EXPERIMENTAL DETAILS

The single crystals of $\beta$–Bi$_2$Pd have been grown in Madrid and Oxford via melt-growth technique from high-purity Bi and Pd in quartz ampoules sealed at 140 mbar He gas. The details are given elsewhere [15]. The resulting single crystals have very similar residual resistivity close to 30 μΩcm, the residual resistivity ratio $RRR$ ~ 2.8 and $T_c \approx 5$ K. For the measurements we used the crystals with a lateral size less than 1 mm and the width ~ 25 μm.

All the measurements have been realized in the Centre of Low Temperature Physics Kosice. Heat capacity measurements have been performed using an ac technique [16,17]. ac calorimetry consists of applying periodically modulated power and measuring the resulting sinusoidal temperature response. In our case, the heat was supplied to the sample at a frequency $\omega$ ~ several Hz by a light emitting diode via an optical fiber. The temperature oscillations were recorded by the chromel-constantan thermocouple calibrated in the magnetic field using measurements on ultrapure silicon. Although ac calorimetry is not capable of measuring the absolute values of the heat capacity, it is very sensitive to relative changes in minute samples and it enables continuous measurements. We performed measurements at temperatures down to 0.55 K in the $^3$He refrigerator and in a horizontal superconducting magnet allowing for a precise angular dependence.

The local magnetometry experiments were performed using a sensor comprising an array of rigid Hall probes based on semiconductor heterostructures with a two-dimensional electron gas as the active layer. The array, prepared in Bratislava, consists of 8 Hall probes arranged in a line; the size of an individual probe is 10 x 10 μm$^2$ with the pitch 25 μm. The sample was placed on top of the sensor and mounted inside a $^3$He cryostat; the magnetic field was applied perpendicular to the $ab$ plane of the crystal via a superconducting coil installed in the cryostat. The sensor was supplied with a constant bias current and the Hall voltage measured across the sensor was directly proportional to perpendicular component of the local magnetic induction as the sample was exposed to an external field. Prior to each measurement the sample was cooled in zero field. Then, gradually increasing magnetic field was applied and sensor response was recorded. When the sample is in ideal diamagnetic state, no magnetic field penetrates the crystal, thus the sensor located below the sample is shielded and no Hall voltage is detected. However, in most of the measurements small initial slope appears as a consequence of finite distance between the probe and the sample, when the probe reads a small portion of unshielded applied field. Note, that this initial slope is removed before further data treatment. When applied field reaches a critical value, called the first penetration field $H_p$, first vortex enters the sample and more of them follow when the applied field is increased further. Thus, when the field rises above the critical value, the Hall voltage of the sensor at the position where the first vortex settles starts to increase.

The transport measurements were done by four probes contacted by a silver paint to the sample which was then placed to $^4$He and $^3$He cryostats and a magnetic field parallel and perpendicular to the basal plane of the sample was applied.

The scanning tunneling spectroscopy experiments were performed by means of a homemade STM head inserted in a commercial Janis SSV cryomagnetic system with a $^3$He refrigerator and controlled by the Nanotec's Dulcinea SPM electronics [18]. In order to obtain a clean sample surface, the sample was cleaved shortly before inserting into the refrigerator and the system was pumped to a high vacuum to minimize the contamination of the sample surface. Prior to measurement, the Au tip was prepared in situ by repetitive impaling into the bulk Au sample and subsequent slow retraction, while recording the current as a function of the tip position at a constant bias voltage. The procedure was repeated until the current–position dependence exhibited clear steps, indicating the conductance quantization and single atom contact phenomena typical for gold. The tip was then scanned over the sample. Bias voltage was applied to the tip, while the sample was grounded; the initial tunneling resistance was set to 1 MΩ.

## III. RESULTS AND DISSCUSSION

### A. ac calorimetry

Figure 1 depicts the heat capacity of the sample and related parameters resulting from measurements at 0 and 1 T while sweeping the temperature. The main panel shows total heat capacity $C/T$ of the sample plus addenda in superconducting ($C_s$) and in normal ($C_n$) states. In the zero field measurement a clear superconducting anomaly is observed close to 5 K, whereas in 1 T superconductivity is completely suppressed. Thus this field measurement may be used to subtract a heat capacity background coming from the lattice and the normal state electronic contributions and addenda as well. Electronic contribution in superconducting state $C_{es}$ may be calculated as $C_{es}/\gamma_n T = \Delta C/\gamma_n T + 1$, where $\Delta C = C_s - C_n$ and $\gamma_n = C(H=1\text{ T})/T|_{T\sim 0K} - C(0\text{ T})/T|_{T\sim 0K}$. The resulting electronic heat capacity (shown in the lower inset of Fig. 1) was compared with the so-called α-model [19] based on the BCS theory. The only parameter in this model is the gap ratio $2\Delta/k_B T_c$. For our measurements we found a very good agreement with the theoretical curve corresponding to a single gap with the coupling ratio ratio $2\Delta/k_B T_c = 4.1$ (blue line in the inset).

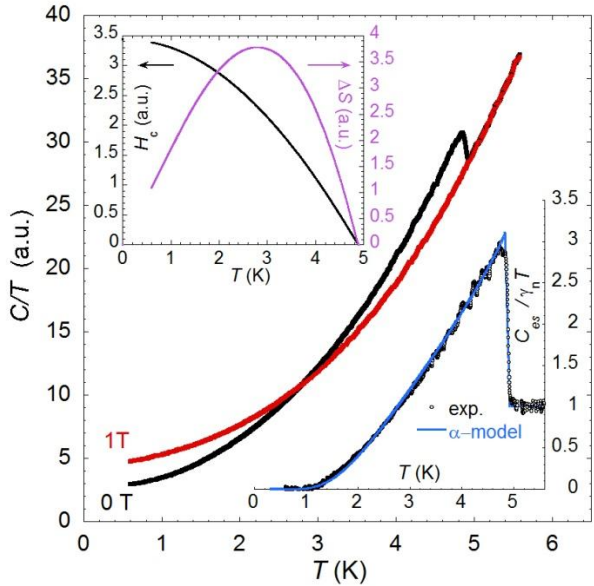

Fig.1 Total heat capacity of the sample plus addenda in superconducting (0 T) and normal (1 T) states. Upper inset: Difference in entropy (purple line, right axis applies) and the thermodynamic critical field (black line, left axis). Lower inset: Electronic specific heat from experimental data and theoretical curve from alpha-model with the coupling ratio $2\Delta/k_B T_c = 4.1$.

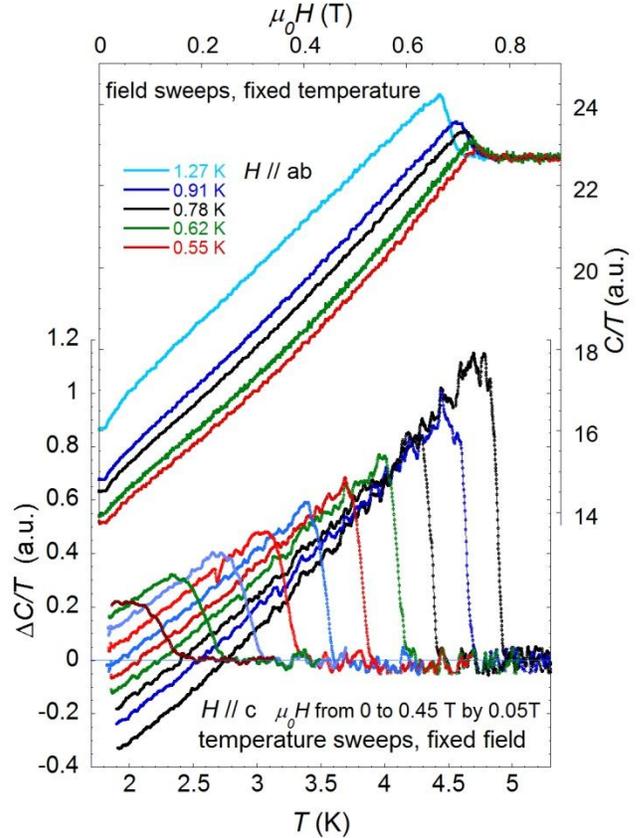

Fig.2 Heat capacity of Bi₂Pd. Field and temperature sweeps at specified fixed temperature and field, resp.

The fit reproduces very well the jump at the anomaly as well as the overall shape of the measured curve. It is obvious that in this data there is no sign of any additional hump or other manifestation of expected second energy gap.

Upper inset of the figure shows a difference in entropy between the superconducting and normal state $\Delta S$ obtained after integration of the $\Delta C/T$ curve. From the second integration of the data we obtain a temperature dependence of the thermodynamic critical field $H_c$. Since the results of ac calorimetry measurements are in arbitrary units, such a calculated $H_c$ is also in arbitrary units. Still it bears information about the coupling strength in the system. The ratio $2[T/H_c(0)](dH_c/dT)|_{T\to T_c}$ is equal to $2\Delta/k_B T_c$ [20] and gives consistently the value 4.1.

Figure 2 combines measurements at fixed temperature while sweeping the magnetic field for $H\|ab$ and those at fixed fields sweeping the temperature for $H\|c$. Temperature sweeps in lower part of the figure show gradual suppression of the superconducting transition temperature with increasing magnetic field, the height of the anomaly diminishes as well. On the other hand, the anomaly in field sweeps (upper part of the figure) is augmented with increasing temperature while moving towards lower fields as expected. The field sweep at the lowest temperature $T = 0.55$ K (the right-most curve) is actually close to the Sommerfeld coefficient $\gamma = \Delta C/T|_{T\to 0}$ which is related to the density of states. It does not show any nontrivial features that could be attributed to the second energy gap. Note that nonlinearity at very low fields is related to the lower critical field $H_{c1}$ of the sample, with complete expulsion of the field from its interior below that value, leading to a constant value of $\Delta C/T$. At 0.55 K $H_{c1}^{ab} \sim 150$ G can be estimated in a fair agreement with the Hall probe magnetometry determination shown below.

From the temperature and field sweeps taken at the both principal field orientations (some shown in Fig. 2) the values of the upper critical field $H_{c2}$ were extracted defined in the midpoint of each anomaly. Resulting temperature dependences, $H_{c2}^{c}$ and $H_{c2}^{ab}$ (symbols), are depicted in Fig. 3 revealing a linear behavior close to the critical temperature and a saturation at lower temperatures. The both curves can be fully described in the framework of the standard Werthamer-Helfand-Hohenberg (WHH) theory [21] in the overall temperature range down to very low temperatures. Lines in the figure represent the respective fit for each $H_{c2}$ curve. Only a small anisotropy $\Gamma = H_{c2}^{ab}/H_{c2}^{c} = 1.25$ is found. Values of $H_{c2}(0)$ as well as the coherence length $\xi$ for both field orientations and superconducting anisotropy are summarized in Table I.

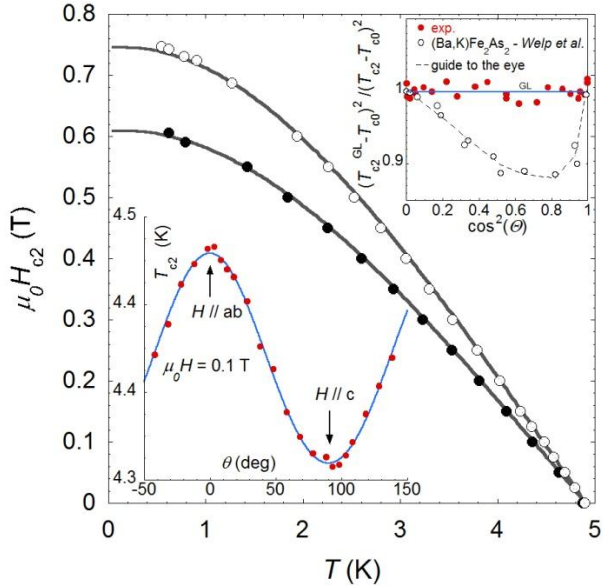

Fig.3 Temperature dependence of the upper critical field for $H$ in the c direction (closed symbols) and parallel to the ab planes (open symbols). Lines are fit to the WHH model. Lower inset: angular dependence of $T_{c2}$ measured at fixed field 0.1 T (symbols) and theoretical curve from Ginzburg-Landau model (line). Upper inset: deviation function of our measurements (red symbols) compared to those of Welp et al. [22] (open circles) for a two gap superconductor, (Ba,K)Fe₂As₂.

In order to inspect closely the evolution of $H_{c2}$ in-between the two main crystallographic orientations we have also performed the heat capacity measurements in magnetic fields oriented at different angles with respect to the ab plane of the sample. Lower inset of Fig.3 shows the angular dependence of the critical temperature $T_{c2}$ at fixed field 0.1 T (symbols). Line is a theoretical curve from the effective mass model within the Ginzburg-Landau theory,

$$T_{c2}(\theta) = T_{c0} + H\sqrt{(\cos^2(\theta) + \Gamma^2 \sin^2(\theta))}/(\partial H_{c2}^{ab}/\partial T), \quad (1)$$

where $T_{c0}$ is the zero-field transition temperature. As can be seen in the inset, formula describes the data perfectly. The quantitative agreement is emphasized in the upper inset of the figure. There, the deviation function is plotted by the closed symbols, showing no difference between the data and the theory. For comparison, the deviation function of a (Ba,K)Fe₂As₂ sample [22] is displayed showing a typical feature related to the two-gap superconductivity. Thus, our heat capacity measurements clearly show the single s-wave gap superconductivity in the system.

### B. Hall-probe magnetometry & STS

Lower critical magnetic field $H_{c1}$ was inspected via local magnetometry measurements. First penetration field $H_p$, taken as a value of the applied field at which the Hall voltage starts to deviate from zero, is related to $H_{c1}$ via demagnetization coefficient $a$, a constant factor depending only on the sample geometry, such that $H_{c1}(T) = aH_p(T)$.

Specifically, $H_{c1} = H_p/\tanh\sqrt{\alpha\frac{d}{2w}}$, where $d$ and $2w$ is the sample thickness and the width, resp. and $\alpha$ equals to 0.36 or 0.67 for strip and disk, resp. [23]. We estimated a value of the factor $a \sim 4.5$ (in between those for disk and strip) corresponding to the width to thickness ratio of the sample being $\sim 8$ for the magnetic field oriented perpendicularly to the sample's basal plane. In the Fig. 4b) magnetic induction vs. applied field, B-H loops measured at indicated temperatures on a probe located close to the sample edge (this position was chosen due to relatively strong pinning [24]) are displayed. In the loops the applied field was first gradually increasing and subsequently decreasing back to zero. For the sake of clarity, only selected loops are shown. Taking the value of $H_p$ from all B-H loops at different temperatures enabled us to construct a complete temperature dependence of $H_{c1}$. The resulting data depicted in Fig. 4a) by blue squares (to be in scale the data is multiplied by a factor of 10) shows a linear behavior close to $T_c$ and a saturation at low temperatures, leading to the value $\mu_0 H_{c1}^{c}(0) = 22.5 \pm 2.5$ mT. The value of the penetration depth $\lambda_{ab}(0)$ corresponding to this field can be extracted from the formula $H_{c1}^{c} = \frac{\phi_0}{4\pi\lambda_{ab}^2}(\ln\kappa_c + \alpha)$, with the Ginzburg-Landau parameter $\kappa_c$ calculated from a relation $\frac{H_{c2}^c}{H_{c1}^c} = \frac{2\kappa_c^2}{\ln\kappa_c + \alpha}$, where $\alpha$ is a parameter close to 0.5. Values of $\lambda_{ab}(0)$ and $\kappa_c$ are listed in Table I. There, also a value of $\lambda_c^*(0)$ is introduced, assuming a unique superconducting anisotropy constant $\Gamma_\lambda = \lambda_c/\lambda_{ab} = \xi_{ab}/\xi_c = \Gamma_\xi = \Gamma_{Hc2}$, usual for a single-gap superconductor. Anisotropy of $H_{c1}$ may be derived from that of $\lambda$ as $\Gamma_\lambda = \Gamma_{H_{c1}}\{1 + \ln\Gamma_\lambda/(\ln\kappa + \alpha)\}$ and subsequently $H_{c1}^{ab}*(0)$ may be calculated as well.

Next, superconducting energy gap of the system can be addressed as we compare the data with a single gap formula

$$H_{c1}(T)/H_{c1}(0) \propto \lambda^{-2}(T)/\lambda^{-2}(0) = 1 - 2\int_{\Delta(T)}^{\infty} \frac{\partial f}{\partial E} \frac{E}{\sqrt{E^2 - \Delta^2(T)}} dE. \quad (2)$$

A very good fit is found (blue line) for the coupling strength of $2\Delta/k_BT_c = 4.1$, consistently with the heat capacity measurements.

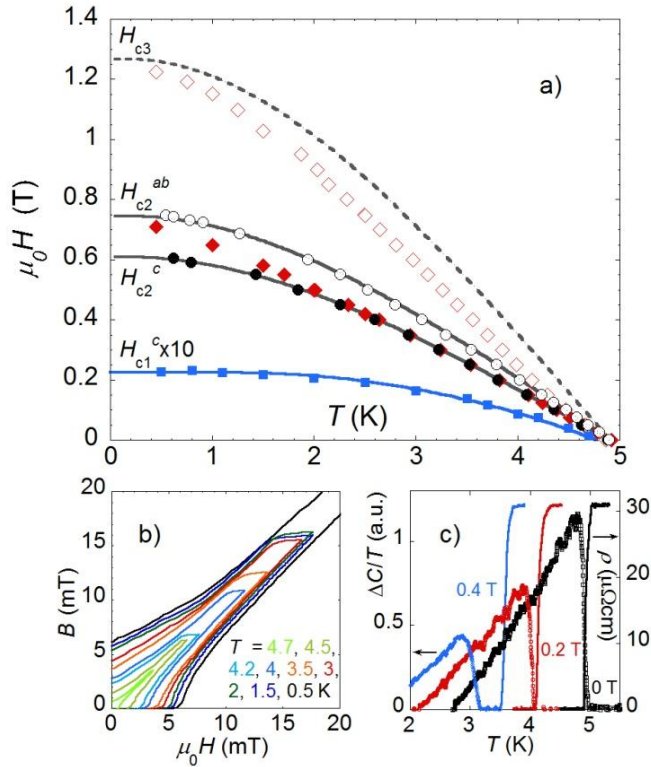

Fig.4 a) The phase diagram of $\beta$-Bi$_2$Pd showing $H_{c1}^c$ (blue squares, the data are multiplied by factor 10 to be in scale) with the best fit (blue line) to Eq. 2 corresponding to $2\Delta/k_BT_c = 4.1$; $H_{c2}$ from heat capacity (circles) and resistivity (lozenges) for both $H\|c$ (filled symbols) and $H\|ab$ (open symbols) together with the WHH theoretical curves (grey full lines) as well as a $H_{c3}$ line (grey dashed line). b) Local magnetic induction $B$ as a function of applied field $H$ measured by a Hall-probe at selected temperatures, with the field first increasing and then decreasing back to zero. c) Superconducting transitions for $H//ab$ revealed by the heat capacity (left axis) and resistivity measurements (right axis) at magnetic fields 0, 0.2 and 0.4 T, respectively.

Hence, both ac calorimetry and Hall-probe magnetometry provide a support for standard single $s$-wave gap superconductivity with the strong coupling ratio in $\beta$-Bi$_2$Pd. We underline that this conclusion has been drawn by the analysis of the data obtained on several samples grown in two different laboratories. A single superconducting gap has also been found by scanning tunneling spectroscopy measurements in our previous paper [15]. There the s-wave superconducting gap with $\Delta = 0.76$ meV has been obtained in the Madrid STM setup placed in a dilution refrigerator. To check the consistency we have performed the STM measurements in the Kosice STM setup placed in a $^3$He refrigerator with a basal sample temperature of 300 mK on the same samples as used for the ac calorimetry and the Hall-probe magnetometry. The obtained BCS-like spectra with a single superconducting gap are in full agreement with the previous measurements [15]. Thus, both the bulk and the surface of $\beta$-Bi$_2$Pd feature a standard BCS-like superconductivity. The only systematic difference is the coupling strength $2\Delta/k_BT_c = 3.7 \pm 0.1$ in both STM measurements, which is significantly smaller than the calorimetric and magnetometric value of $2\Delta/k_BT_c = 4.1$. Yet, it indicates certain differences between the bulk and surface properties of $\beta$-Bi$_2$Pd, respectively. Smaller coupling strength could be caused by a proximity effect from bulk to a metallic surface. Other possibilities are discussed below.

### C. Magnetotransport

Magnetotransport measurements have been performed on several crystals as well. The temperature dependence of the electrical resistivity $\rho(T)$ with a convex curvature above 50 K and the residual resistivity ratio $RRR \sim 2.8$ reproduces well the behavior presented in Fig. 1 of our previous paper [15]. The superconducting transition temperature in zero field in different crystals varies less than 0.1 K being $T_c = 4.93$ K at the 50% of the transition for the crystal shown in the Fig. 4c) (black line, right axis). The width of the transition defined between 10% and 90% of the transition is less than 0.1 K.

Fig. 4c) shows also the jump in the heat capacity of the same sample at zero magnetic field (black symbols, left axis) with the onset of the transition close to the zero resistance edge. The midpoint of the jump is at 4.89 K and the width between 10 and 90% of the jump is 70 mK pointing to extreme homogeneity of the sample. The other crystals show the same narrow transitions and $T_c$ between 4.9 and 5.0 K. Figure 4a) shows by lozenges the upper critical fields parallel and perpendicular to the basal plane obtained from the resistive transitions defined at $0.5\times\rho_n$, the normal state resistivity. Also the upper critical fields inferred from the heat capacity are shown by the circles for comparison. For $H//c$ a good agreement is found between the two determinations at $T > 2$ K while at lower temperatures the agreement is lost as the resistively determined $H_{c2}(T)$ displays higher values and does not develop any saturation. Such a deviation from the standard WHH behavior is found in many superconductors and can be explained for example by strong-coupling effects [25]. But we believe that the thermodynamic determination of $H_{c2}$ from a clear second order phase transition as shown in Fig. 2 is much more reliable and provides the true upper critical field values of the bulk. We remark that $H_{c2}^{c}(T)$ from the heat capacity is also in a very good agreement with the data obtained by the magnetic susceptibility measurements [15].

Even bigger discrepancy between the heat capacity and resistively determined data is found in the temperature dependence of the upper critical field $H_{c2}^{ab}(T)$. There, the resistively determined data (open lozenges) completely mismatch the dependence obtained from the heat capacity (open circles). This can also be seen directly in the source data depicted in the inset of Fig. 4c) where for increasing field the onset of the heat capacity anomaly gets more and more below the onset of finite resistivity. The overall temperature dependence of the resistively determined $H_{c2}^{ab}(T)$ is similar to that of the resistive $H_{c2}^{c}(T)$ showing a little negative curvature without a real saturation at low temperatures but for parallel fields even a positive curvature appears above 3 K. It is noteworthy that the data are very close to the supposed third critical field $H_{c3}(T)$ calculated from the thermodynamic upper critical field as $1.695 \times H_{c2}^{ab}(T)$ [26] for the external magnetic field parallel to the sample surface. Indeed, the configuration of the four probes in the resistivity measurements is favorable to detect $H_{c3}$ in our samples, with all the probes put on the surface parallel to the applied field. Determination of absolute value of $H_{c3}$ is somewhat uncertain since the resistivity approaches the normal value in the surface sheath gradually and is also affected by measuring current [27]. In our case extremely small current densities below 0.1 A/cm$^2$ were used for the measurements which is appropriate for $H_{c3}$ determination. Obviously, a more detailed study is needed to firmly determine the origin of the resistive critical fields which beside the surface superconductivity [28,29] could also be affected for example by filamentary phases.

We have noticed that our resistively obtained data are very close to the $H_{c2}^{ab}(T)$ dependence determined by Imai et al. [12]. In their case it was taken as an evidence for the concept of multigap superconductivity in $\beta$-Bi$_2$Pd showing a positive curvature of $H_{c2}^{ab}(T)$ and a temperature dependent superconducting anisotropy $\Gamma(T) = H_{c2}^{ab}/H_{c2}^{c}$, but nothing like this is found in the thermodynamically determined upper critical fields.

The surface superconductivity in $\beta$-Bi$_2$Pd is particularly interesting because the topologically protected surface states have recently been detected by ARPES measurements [14]. It could have implications on superconductivity and feature unconventional superconducting order parameter with gapless states on the surface [30]. But our measurements have proven that both the bulk and surface superconductivity is characterized by a conventional single $s$-wave gap. On the other hand the observed surface effects reflected in the resistively determined critical fields and in the lower value of the superconducting coupling strength $2\Delta/k_BT_c$ as found by the STS measurements ask for further explanation.

## IV. CONCLUSIONS

The comprehensive calorimetric studies via the sensitive ac technique have shown that $\beta$-Bi$_2$Pd is a single s-wave gap superconductor with a strong coupling of $2\Delta/k_BT_c = 4.1$ This conclusion is supported by another bulk probe, the Hall-probe magnetometry but also the subkelvin STS measurements points to the same character of superconductivity. Thus, both the bulk and the surface of the sample are characterized by a conventional superconductivity and no indications for multigap signatures are found. The critical fields, the coherence length and the penetration depth are determined showing a very moderate superconducting anisotropy. The surface sensitive magnetotransport as well as STS measurements indicate that some characteristics differ from their bulk values but if this is connected with the recently discovered topologically protected surface states in $\beta$-Bi$_2$Pd remains for future studies.

## ACKNOWLEDGEMENTS

This work was supported by the projects APVV-0036-11, APVV-14-0605, VEGA 2/0135/13, VEGA 1/0409/15, by the ERDF EU grant under contract No. ITMS26220120005 and the COST action MP1201 as well as by the U.S. Steel Kosice, s.r.o. Sample growth was partially supported by the Spanish MINECO (FIS2014-54498-R), the Region of Madrid Nanofrontmag-CM (S2013/MIT-2850) and Engineering and Physical Sciences Research Council, UK.

TABLE I. Parameters of $\beta$-Bi$_2$Pd superconductor obtained from thermodynamic and magnetometry measurements: lower critical field $H_{c1}$ and its anisotropy $\Gamma_{Hc1}$, penetration depth $\lambda$ and its anisotropy $\Gamma_\lambda$, upper critical field $H_{c2}$ and its anisotropy $\Gamma_{Hc2}$, coherence length $\xi$, and Ginzburg-Landau parameter $\kappa$, all of them for both $H\|c$ and $H\|ab$. The values with asterisk are calculated assuming that $\Gamma_\lambda=\Gamma_\xi$ (see text for details).

| $\mu_0 H_{c1}^c(0)$ | $22.5 \pm 2.5$ mT | $\lambda_{ab}(0)$ | $132 \pm 10$ nm | $\mu_0 H_{c2}^c(0)$ | 0.61 T | $\xi_{ab}(0)$ | $23.2 \pm 0.2$ nm | $\kappa_c = \lambda_{ab}/\xi_{ab}$ | 5.7 |
|---|---|---|---|---|---|---|---|---|---|
| $\mu_0 H_{c1}^{ab*}(0)$ | $19.7 \pm 4.5$ mT | $\lambda_c^*(0)$ | $105 \pm 8$ nm | $\mu_0 H_{c2}^{ab}(0)$ | 0.76 T | $\xi_c(0)$ | $18.5 \pm 0.3$ nm | $\kappa_{ab} = \lambda_{ab}/\xi_c$ | 7.1 |
| $\Gamma_{Hc1}$ | 1.14 | $\Gamma_\lambda$ | 1.25 | $\Gamma_{Hc2}$ | 1.25 | | | | |